\documentclass[twocolumn,prl,showpacs]{revtex4}

\usepackage{graphicx}
\usepackage{dcolumn}
\usepackage{bm}

\begin{document}

\title{
Large quantum nonlinear dynamic susceptibility of single-molecule
magnets
}

\author{
F. Luis, V. Gonz\'alez, A. Mill\'an, and J. L. Garc\'{\i}a-Palacios
}

\address{Instituto de Ciencia de Materiales de Arag\'on,
C.S.I.C.--Universidad de Zaragoza, 50009 Zaragoza, Spain}
\date{\today}

\begin{abstract}
The nonlinear dynamical response of Mn$_{12}$ single-molecule magnets
is experimentally found to be very large, quite insensitive to the
spin-lattice coupling constant, and displaying peaks reversed with
respect to classical superparamagnets. It is shown that these features
are caused by the strong field dependence of the relaxation rate due
to the detuning of energy levels between which tunneling takes
place. The nonlinear susceptibility technique, previously overlooked,
is thus proposed as a privileged probe to ascertain the occurrence of
quantum effects in mesoscopic magnetic systems.
\\
\flushright{Published in {\em Phys. Rev. Lett.} {\bf 92}, 107201
    (2004).}
\end{abstract}

\pacs{75.50.Xx, 75.50.Tt, 75.45.+j, 75.40.Gb}
\maketitle

The possibility to observe quantum phenomena, such as tunneling or
even quantum coherence, occurring in mesoscopic magnetic or
superconducting materials is a subject of considerable interest
\cite{Leggett02}. These systems are also important for the study of
the fuzzy border between the classical and quantum worlds
\cite{Zurek91}. As the size increases, tunneling probabilities
decrease exponentially and the system becomes more entangled with its
environment (phonons, electromagnetic fields, nuclear spins,
electrons, etc). This interaction plays a crucial role as it enables
the system to approach equilibrium \cite{Prokof'ev96} but also causes
decoherence of its quantum dynamics \cite{Leggett85}. Besides its
fundamental interest, decoherence sets the limits for application of
the quantum behavior of mesoscopic systems to, e.g., quantum
computing.

Molecular crystals are ideal for these studies \cite{Gatteschi94}.
They are ordered lattices of identical molecules, each containing a
magnetic cluster that is isolated from its neighbors by a shell of
organic ligands. Because of strong intramolecular super-exchange
interactions, each cluster behaves at low temperatures as a
superparamagnet with a large spin $S$ that flips as a whole.  The best
studied of these is the Mn$_{12}$ acetate, for which $S=10$. Its
strong magnetic anisotropy produces magnetic relaxation and hysteresis
at low temperatures \cite{Sessoli93}, which make it a potential
candidate for information storage at the molecular level. The spin
reversal takes place via quantum tunneling through the anisotropy
energy barrier whenever the magnetic field brings into resonance
magnetic states located at both sides of the anisotropy barrier
\cite{Friedman96}.

For classical superparamagnets, Garc\'{\i}a-Palacios and Svedlindh
\cite{Garcia00} predicted that the nonlinear dynamic susceptibility
can be very large and, in contrast to the linear susceptibility, quite
sensitive to the spin-bath coupling strength (damping parameter). In
this work, we apply this technique, unexploited in this field, to
investigate environmental effects in molecular clusters. We have
studied two compounds, both having the same Mn$_{12}$ magnetic core,
but different ligand molecules. This enables us to investigate how the
spin dynamics is affected by different couplings to the phonon bath.

Our results show that the nonlinear response of these single-molecule
magnets is dominated by a very large contribution, hitherto
unforeseen, whose sign is {\em opposite\/} to the classical one. This
contribution is interpreted in terms of the extreme sensitivity of
quantum tunneling probabilities to the detuning of the magnetic levels
by the external field. Therefore the dynamical nonlinear
susceptibility permits to ascertain unambiguously if tunneling plays a
role in the relaxation of magnetic molecular clusters and
nanoparticles. In addition to its fundamental interest, this
phenomenon might be relevant in situations where nonlinear effects
become important (e.g., magneto-optical applications of these
systems).

The preparation of the Mn$_{12}$ acetate,
[Mn$_{12}$(CH$_{3}$COO)$_{16}$(H$_{2}$O)$_{4}$O$_{12}$]$\cdot
2$CH$_{3}$COOH$\cdot 4$H$_{2}$O, was based on the original method of
Lis \cite{Lis80}. The Mn$_{12}$ $2$-Cl benzoate,
[Mn$_{12}$($2$ClPhCOO)$_{16}$(H$_{2}$O)$_{4}$O$_{12}$]$\cdot$CH$_{2}$Cl$_{2}
\cdot 5$H$_{2}$O,
hereafter abbreviated to Mn$_{12}$Cl, was prepared following the
method described in Ref.~\cite{Ruiz98}. XRD patterns of powders of
both materials show peaks corresponding to the crystal structures
described in the literature \cite{Lis80,Ruiz98}. In order to avoid
uncertainties concerning the orientation of the crystals, we measured
powdered samples embedded in a two-component glue.

The nonlinear susceptibility is usually obtained (see, e.g.,
Ref.~\cite{Bitoh93}) from the third-harmonic of the response to an ac
magnetic field $\sim\Delta h\,{\rm e}^{{\rm i}\omega t}$.
Alternatively \cite{Wu93}, one can consider the first-harmonic
response $\chi(\omega)$ but in the presence of weak superimposed dc
fields $H$, from which one can get a quantity that posseses similar
dependences on the relevant parameters (frequency, damping, etc.).
Specifically, the (first) {\em nonlinear susceptibility\/} $\chi_{3}$
was obtained by fitting the real and imaginary components of the
field-dependent ac susceptibility to a polynomial \cite{fit}
\begin{equation}
\chi(\omega,H)
=
\chi_{1}(\omega)
+
3\chi_{3}(\omega)H^{2}
+
5\chi_{5}(\omega)H^{4}
+
\cdots
\;,
\label{nonlinear}
\end{equation}
where $\chi_{1}$ is the custom linear susceptibility.
The ``expansion coefficients'' $\chi_{1},\chi_{3}, \ldots$ are
$H$-independent and the factors $1,3,5,\ldots$ arise when raising to
$n$ in $\chi_{n}(H+\Delta h)^{n}$ and retaining terms linear in
$\Delta h$.
Below we will show that the classical $\chi_{3}$ obtained in the above
way is equivalent to the third-harmonic susceptibility.

\begin{figure}
\resizebox{8.cm}{!}{\includegraphics{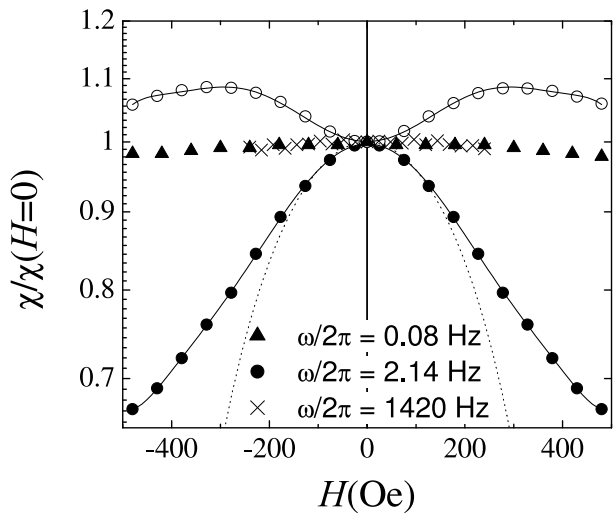}}
\caption{ Dynamical susceptibility of Mn$_{12}$Cl, normalized by its
zero-field value, measured at $T = 4.5$ K as a function of the dc
magnetic field. Full symbols, real part; open symbols, imaginary
part. The frequencies shown correspond to $\omega \tau = 0.015$,
$0.4$, and $265$. The solid lines represent polynomial fits (of order
$8$) from which the non-linear susceptibility is obtained [see
Eq.~(\ref{nonlinear})]. The dotted line is the parabolic term
$\chi_{1} + 3\chi_{3}H^{2}$ that dominates the low-field behavior.}
\label{Fig1}
\end{figure}

All magnetic measurements were performed using a commercial SQUID
magnetometer.
Measurements were performed in a broad range of temperatures $1.8$\,K
$< T < 60$\,K and frequencies $0.025$\,Hz $< \omega/2\pi < 1.5$\,kHz.
We took especial care to ensure that the amplitude of the ac field
($\Delta h=4.5$\,Oe) was sufficiently small not to induce any
additional nonlinearity (i.e., $\chi_{1},\chi_{3}, \ldots$, are
independent of $\Delta h$).
Previous results on $\chi(\omega,H)$ \cite{Hernandez96} contain the
dependences of the linear and nonlinear susceptibilities mixed,
whereas the above method enables one to separate and analyze the
genuine nonlinear contributions.
An example of its application is shown in Fig.~\ref{Fig1}.

\begin{figure}
\resizebox{8.cm}{!}{\includegraphics{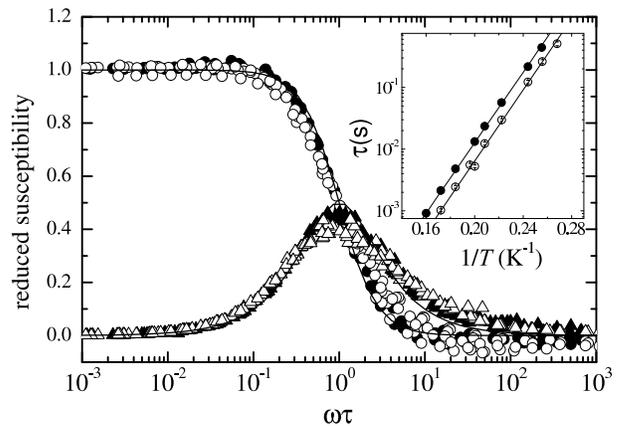}}
\caption{Real (circles) and imaginary (triangles) parts of the reduced
linear susceptibility $(\chi_{1}-\chi_{S})/(\chi_{T}-\chi_{S})$,
measured between $T = 3.7$\,K and $6.6$\,K in the frequency range
$2.5\times 10^{-2}$\,Hz $<\omega/2 \pi < 1.5\times 10^{3}$\,Hz. The
lines are calculated using Debye law [Eq.~(\ref{Debye})]. The
inset shows the temperature dependence of the relaxation time
$\tau$. Full and open symbols correspond to Mn$_{12}$ acetate and
Mn$_{12}$Cl, respectively.} \label{Fig2}
\end{figure}

{\em Linear\/} ac susceptibility data measured as a function of
$\omega$ at zero field are shown in Fig.~\ref{Fig2}. The results are
well described by a Debye law
\begin{equation}
\chi = \chi_{S} + \frac{\chi_{T}-\chi_{S}}{1+ {\rm i}\,
\omega\tau} \;, \label{Debye}
\end{equation}
where $\chi_{T}$ and $\chi_{S}$ are the isothermal (equilibrium) and
adiabatic limits of $\chi_{1}$. The samples have therefore
well-defined relaxation times $\tau$, which follow an Arrhenius law
$\tau=\tau_{0} \exp{(U/k_{\rm B}T)}$. The activation energy $U$ is the
same for both compounds (same slopes in the inset of Fig.~\ref{Fig2}),
i.e., tunneling takes place via the same excited levels, corresponding
approximately to the tunnel-split $m=\pm 4$ states. In contrast, the
pre-exponential factor $\tau_{0}$ of Mn$_{12}$Cl is a factor two
shorter than $\tau_{0}$ of Mn$_{12}$ acetate. $\tau_{0}$ is related to
the lifetime of excited states, which decay by emitting phonons to the
lattice.  Therefore, the spin-lattice interaction, or damping, appears
to be stronger for Mn$_{12}$Cl, in agreement with previous results
\cite{Luis99}. This shows that these two materials are suited to study
how a magnetic cluster behaves in different environments.

\begin{figure}
\resizebox{8.cm}{!}{\includegraphics{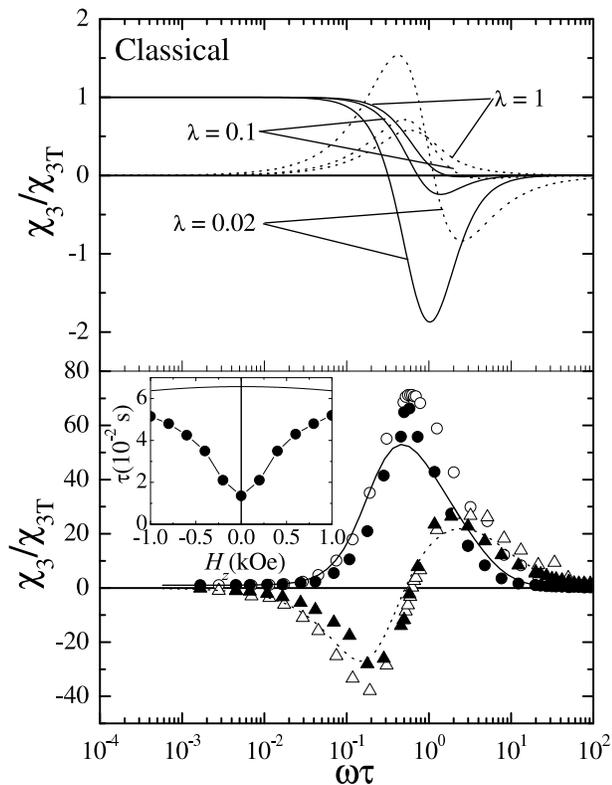}}
\caption{Upper panel: nonlinear susceptibility of classical spins,
calculated by the method of Ref.~\cite{Garcia00}, at $k_{\rm B}T/U =
5/70$ for three damping coefficients $\lambda$.  Full lines, real
part; dotted lines, imaginary part. Lower panel: nonlinear
susceptibility of Mn$_{12}$ acetate (full symbols) and Mn$_{12}$Cl
(open symbols) measured at $T=5$\,K. Dots and triangles represent the
real and imaginary parts. We have multiplied $\omega$ by the zero
field $\tau=1.31\times 10^{-2}$\,s and $5.55 \times 10^{-3}$\,s, for
Mn$_{12}$ acetate and Mn$_{12}$Cl, respectively. The lines are quantum
calculations performed as in Ref.~\cite{Luis98}: full line, real part;
dotted line, imaginary part.  The inset shows the relaxation time of a
single crystal of Mn$_{12}$ acetate measured with the field applied
along the anisotropy axis ($\bullet$) and the classical result (full
line).}
\label{Fig3}
\end{figure}

Before presenting the results for the {\em nonlinear\/} susceptibility
for these systems, let us consider the theoretical behavior of this
quantity in the {\em classical\/} limit. We have calculated $\chi_{3}$
from the exact (continued-fraction) solution of the corresponding
Fokker--Planck equation \cite{Garcia00,Kalmykov97}. To mimic the
experiments: (i) the calculation is done for spins with randomly
distributed anisotropy axes and (ii) the first harmonic of the
response is computed in weak dc fields, and the quadratic coefficient
extracted [see Eq.~(\ref{nonlinear})].  The results are displayed in
Fig.~\ref{Fig3} for various values of the Landau--Lifshitz damping
parameter $\lambda$. It is seen that the $\omega$ dependence of the
so-obtained $\chi_{3}$, as well as its damping sensitivity, are
qualitatively similar to those of the third harmonic $\chi_{3}$
studied in Ref.~\cite{Garcia00}. In the classical case $\lambda$
determines the time available for the spin to precess around the
anisotropy axis, before decaying to the bottom of the potential wells.

The experimental $\chi_{3}$ data of both compounds are also shown in
Fig.~\ref{Fig3} (lower panel). They agree, when $\omega\tau\ll1$
(sufficiently high temperatures or low frequencies), with the expected
equilibrium behavior $\chi_{3T} \propto 1/T^{3}$. When
$\omega\tau\gtrsim1$, $\chi_{3}$ becomes strongly dependent on
$\omega$. It exhibits a large maximum $\chi^{\prime}_{3} \simeq 60
\chi_{3T}$ near $\omega\tau=1$, which depends weakly on the damping
(i.e., on $\tau_{0}$). However, even more striking is the fact that
the experimental and classically calculated peaks of
$\chi_{3}(\omega)$ have {\em opposite} signs!

We next try to explain the origin of this large and (classically)
unexpected nonlinear response. We observe that Eq.~(\ref{Debye}) also
describes the susceptibility of Mn$_{12}$ in moderate
fields. Therefore, $\chi_{3}$ can be estimated by differentiating
Debye's expression with respect to $H$, getting
\begin{equation}
\chi_{3} \sim \frac{\chi_{3T}}{1+{\rm i}\,\omega \tau}
-
\frac
{{\rm i}\, \omega \chi_{T}}
{6 \left( 1+{\rm i}\, \omega \tau\right)^{2}}
\frac{\partial^{2} \tau}{\partial H^{2}}
\Big|_{0}
\;,
\label{Debyenl}
\end{equation}
where the subscript indicates evaluation at $H=0$. The second,
dominant, term is directly related to the field derivative of the
relaxation time. For a classical spin, the main contribution comes
from the derivative with respect to the component of the field that is
perpendicular to the anisotropy axis \cite{Garcia00}. The effect of
the longitudinal field $H_{z}$ is expected to be small. The reason is
that, classically, the energy scale over which $\tau$ changes is
determined by the very large anisotropy field $H_{K} \simeq
10^{5}$\,Oe. Note also that, classically, $\partial^{2}\tau/\partial
H^{2}|_{0} < 0$ (see the inset of Fig.~\ref{Fig3}).

For a molecular cluster, however, the magnetization reversal takes
place by quantum tunneling. As in a two-level problem, the probability
of tunneling is determined by the ratio between the tunnel splitting
$\Delta$ and the bias $\xi$ generated by $H_{z}$. For states with
$m=\pm 4$ of Mn$_{12}$ we have $\Delta\sim 10^{-1}$\,K. As a result,
$\tau$ {{\em increases\/} sharply between $H_{z}=0$ and a few hundred
Oe ($\ll H_{K}$), as the inset of Fig.~\ref{Fig3} shows
\cite{Friedman96}. The large ``positive'' contribution to $\chi_{3}$
that we observe near $\omega \tau = 1$ can therefore be associated to
the strong dependence of the tunneling probability on $H_{z}$, which
renders $\partial^{2}\tau/\partial H^{2}|_{0} > 0$.

In Fig.~\ref{Fig3}, the experimental data are also compared with
theoretical calculations that take into account the quantum nature of
the molecular spins \cite{Luis98,calculations}. The relaxation time
and susceptibility are calculated by solving a standard Pauli master
equation, which includes transition rates between energy eigenstates
induced by the interaction with the phonon bath.  Emulating the
experimental conditions, we have computed $\chi^{\prime}$ and
$\chi^{\prime \prime}$ for different applied fields $|H| > 50$\,Oe and
fitted them to Eq.~(\ref{nonlinear}).  Despite the fact that the
calculation does not take into account the precession of the spins,
which produces the main classical contribution, it accounts remarkably
well for the experimental results, including the change of sign with
respect to the classical prediction.

Before concluding, let us briefly consider the limit of large dc
fields. Then the relaxation time approaches its classical limit,
because the field blocks tunneling via all levels lying in energy
below the classical anisotropy barrier. In order to check how this
change of the relaxation process modifies the nonlinear response, we
have extracted, using the same method described above, $\chi_{3}$ from
experimental data measured at $ |H| > 1.5 \times 10^{3}$\,Oe. The
results, plotted in Fig.~\ref{Fig4}, resemble indeed the predictions
for classical spins. In particular, the inequality
$|\chi_{3}^{\prime}|\leq|\chi_{3T}|$ is fulfilled in the whole frequency
range.
\begin{figure}[b]
\resizebox{8.cm}{!}{\includegraphics{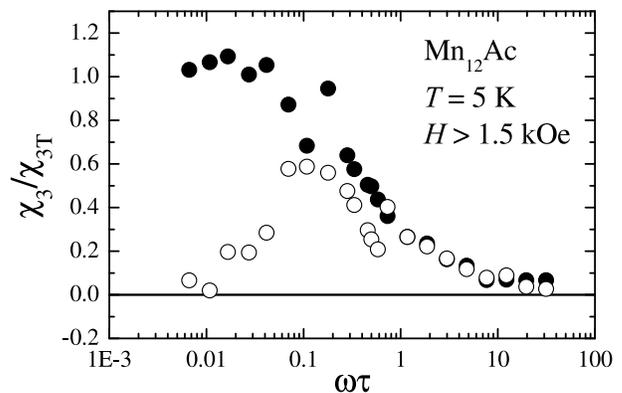}}
\caption{Real ($\bullet$) and imaginary ($\circ$) nonlinear
susceptibilities of Mn$_{12}$ acetate obtained as described in the
text [see Eq.~(\ref{nonlinear})], from susceptibility data
measured for $ |H| \geq 1.5 \times 10^{3}$ Oe.}
\label{Fig4}
\end{figure}

Interestingly, these high-field results agree well with
calculations for overdamped classical spins ($\lambda \gtrsim 1$
in the upper panel of Fig.~\ref{Fig3}). This is at first
surprising since classically $\lambda \approx 2 \pi /\omega_{L}
\tau_{0}$, where $\omega_{L} = g \mu_{B} H_{K} /\hbar$ is the
Larmor frequency. For Mn$_{12}$ clusters, $2 \pi/ \omega_{L} \sim
4 \times 10^{-12}$\,s while $\tau_{0} \simeq 1.3 \times
10^{-8}$\,s and $2.7\times10^{-8}$\,s for Mn$_{12}$Cl and
Mn$_{12}$ acetate, respectively. This gives $\lambda \sim
10^{-4}$. For such a low damping the classical contribution is
expected to dominate over the quantum one, whereas clearly the
opposite is observed.

We argue, however, that the situation might become different if
the spins relax via quantum tunneling. Their precession around the
anisotropy axis, as well as any other coherent dynamics, are then
limited not by the decay towards lower lying energy states, but
rather by the tunneling time $\tau_{\rm tun} \equiv \hbar/\Delta$
\cite{Luis98}. For times longer than $\tau_{\rm tun}$ the system
evolves towards an incoherent mixture of delocalized (i.e., with
$\langle S_{z}\rangle \sim 0$) energy eigenstates. Near the top of
the energy barrier (i.e., $|m| \leq 4$), we have $\tau_{\rm tun}
\lesssim 5 \times 10^{-11}$\,s. Replacing $\tau_{\rm tun}$ for
$\tau_{0}$ in the above $\lambda$ gives $\lambda \sim 0.1$,
approximately equal for both compounds. Under these conditions,
$\chi_{3}$ is expected to depend weakly on the spin-lattice
interaction, which agrees with the observed features. In addition,
our results show that coherence is lost for times much shorter
than $\tau_{0}$.

In summary, we have investigated experimentally the nonlinear
dynamical susceptibility of molecular nanomagnets and performed
theoretical calculations at both the classical and quantum levels of
description. We have found that Mn$_{12}$ single-molecule magnets show
a large quantum nonlinear magnetic response, which can be directly
linked to the tunneling mechanism by which their spins flip, and whose
sign is opposite to the classical predictions. This phenomenon should
not be restricted to Mn$_{12}$, but it could be observable as well for
other mesoscopic magnetic systems (e.g., other magnetic clusters and
nanoparticles).  The nonlinear magnetic susceptibility therefore
offers a privileged probe to ascertain if tunneling plays a role in
these mesoscopic materials, for which standard (in particular, linear)
relaxation experiments do not always provide a definite answer
\cite{Mamiya02}.

The authors benefitted from enlightening discussions with J. F.
Fern\'andez and P. Svedlindh. This work has been funded by grants
MAT$02-0166$ and BFM$2002-00113$ from the Spanish Ministerio de
Ciencia y Tecnolog\'{\i}a. V.G. acknowledges a grant from Consejo
Superior de Investigaciones Cient\'{\i}ficas.


\begin{thebibliography}{99}
\bibitem{fit} The fit was done
using polynomials of increasing order until convergence of
$\chi_{3}$ was observed. The results agree with those obtained by
fitting only the susceptibility in the low field region (parabolic
dependence, see Fig.~\ref{Fig1}).
\bibitem{calculations} The field-dependent ac susceptibility was
calculated as described in Ref.~\cite{Luis98} averaging over a random
orientation of easy axes and over a Gaussian distribution of bias
dipolar fields. Its width $\sigma = 190$\,Oe was estimated from $2 g
\mu_{B} S \sigma = k_{\rm B} T_{c}$, where $T_{c} \sim 0.5$\,K is the
ordering temperature extracted from Monte Carlo simulations
\cite{Fernandez00}.
\end{thebibliography}
\end{document}